\input harvmac
\input epsf
\input amssym
%
%
\noblackbox
\newcount\figno
\figno=0
\def\fig#1#2#3{
\par\begingroup\parindent=0pt\leftskip=1cm\rightskip=1cm\parindent=0pt
\baselineskip=11pt
\global\advance\figno by 1
\midinsert
\epsfxsize=#3
\centerline{\epsfbox{#2}}
\vskip -21pt
{\bf Fig.\ \the\figno: } #1\par
\endinsert\endgroup\par
}
\def\figlabel#1{\xdef#1{\the\figno}}
\def\encadremath#1{\vbox{\hrule\hbox{\vrule\kern8pt\vbox{\kern8pt
\hbox{$\displaystyle #1$}\kern8pt}
\kern8pt\vrule}\hrule}}

\def\frac#1#2{{#1 \over #2}}

\def\semi{\subset\kern-1em\times\;}
\def\bar#1{\overline{#1}}
\def\sqr#1#2{{\vcenter{\vbox{\hrule height.#2pt
\hbox{\vrule width.#2pt height#1pt \kern#1pt \vrule width.#2pt}
\hrule height.#2pt}}}}

\def\T{{\bf T}}

\def\ad{\bar a}

%

%

\def\ap{\alpha'}
\def\Nb{\overline{N}}
\def\Qb{\overline{Q}}

\def\xt{\tilde{x}}
\def\psit{\tilde{\psi}}
\def\phit{\tilde{\phi}}
\def\yt{\tilde{y}}
\def\T{{\rm tube}}
\def\B{{\rm BMPV}}

\def\IR{\Bbb{R}}

\lref\myers{ R.~C.~Myers,
``Dielectric-branes,''
JHEP {\bf 9912}, 022 (1999)
[arXiv:hep-th/9910053].
 }
\lref\supertube{
D.~Mateos and P.~K.~Townsend,
``Supertubes,''
Phys.\ Rev.\ Lett.\  {\bf 87}, 011602 (2001)
[arXiv:hep-th/0103030].
}

\lref\mathurlunin{
O.~Lunin and S.~D.~Mathur,
``Metric of the multiply wound rotating string,''
Nucl.\ Phys.\ B {\bf 610}, 49 (2001)
[arXiv:hep-th/0105136].
}

\lref\LuninJY{
O.~Lunin and S.~D.~Mathur,
``AdS/CFT duality and the black hole information paradox,''
Nucl.\ Phys.\ B {\bf 623}, 342 (2002)
[arXiv:hep-th/0109154].
}

\lref\mathurstretch{
O.~Lunin and S.~D.~Mathur,
 ``Statistical interpretation of Bekenstein entropy for systems with a
stretched horizon,''
Phys.\ Rev.\ Lett.\  {\bf 88}, 211303 (2002)
[arXiv:hep-th/0202072].
}

\lref\LuninBJ{
O.~Lunin, S.~D.~Mathur and A.~Saxena,
``What is the gravity dual of a chiral primary?,''
Nucl.\ Phys.\ B {\bf 655}, 185 (2003)
[arXiv:hep-th/0211292].
}

\lref\LuninIZ{
O.~Lunin, J.~Maldacena and L.~Maoz,
``Gravity solutions for the D1-D5 system with angular momentum,''
arXiv:hep-th/0212210.
}

\lref\mathur{
S.~D.~Mathur, A.~Saxena and Y.~K.~Srivastava,
``Constructing 'hair' for the three charge hole,''
arXiv:hep-th/0311092.
}

\lref\MathurSV{
S.~D.~Mathur,
``Where are the states of a black hole?,''
arXiv:hep-th/0401115.
}

\lref\ifd{
I.~Bena,
``The polarization of F1 strings into D2 branes: 'Aut Caesar aut nihil',''
Phys.\ Rev.\ D {\bf 67}, 026004 (2003)
[arXiv:hep-th/0111156].
}
\lref\nbi{
A.~A.~Tseytlin,
``On non-abelian generalisation of the Born-Infeld action in string  theory,''
Nucl.\ Phys.\ B {\bf 501}, 41 (1997)
[arXiv:hep-th/9701125].
}
\lref\HerdeiroAP{
C.~A.~R.~Herdeiro,
``Special properties of five dimensional BPS rotating black holes,''
Nucl.\ Phys.\ B {\bf 582}, 363 (2000)
[arXiv:hep-th/0003063].
}
\lref\unp{W.~I.~Taylor, unpublished}
\lref\wati{
W.~I.~Taylor,
``Adhering 0-branes to 6-branes and 8-branes,''
Nucl.\ Phys.\ B {\bf 508}, 122 (1997)
[arXiv:hep-th/9705116].
}
\lref\tvr{
W.~I.~Taylor and M.~Van Raamsdonk,
``Multiple Dp-branes in weak background fields,''
Nucl.\ Phys.\ B {\bf 573}, 703 (2000)
[arXiv:hep-th/9910052].
W.~I.~Taylor and M.~Van Raamsdonk,
``Multiple D0-branes in weakly curved backgrounds,''
Nucl.\ Phys.\ B {\bf 558}, 63 (1999)
[arXiv:hep-th/9904095].
}
\lref\bak{
D.~Bak and K.~M.~Lee,
``Noncommutative supersymmetric tubes,''
Phys.\ Lett.\ B {\bf 509}, 168 (2001)
[arXiv:hep-th/0103148].
}
\lref\emparan{
R.~Emparan, D.~Mateos and P.~K.~Townsend,
``Supergravity supertubes,''
JHEP {\bf 0107}, 011 (2001)
[arXiv:hep-th/0106012].
}
\lref\bmpv{
J.~C.~Breckenridge, R.~C.~Myers, A.~W.~Peet and C.~Vafa,
``D-branes and spinning black holes,''
Phys.\ Lett.\ B {\bf 391}, 93 (1997)
[arXiv:hep-th/9602065].
}

\lref\HorowitzNW{ G.~T.~Horowitz and J.~Polchinski,
``A correspondence principle for black holes and strings,''
Phys.\ Rev.\ D {\bf 55}, 6189 (1997) [arXiv:hep-th/9612146].
}

\lref\BakKQ{
D.~Bak and K.~M.~Lee,
``Noncommutative supersymmetric tubes,''
Phys.\ Lett.\ B {\bf 509}, 168 (2001)
[arXiv:hep-th/0103148].
}

\lref\EmparanUX{
R.~Emparan, D.~Mateos and P.~K.~Townsend,
``Supergravity supertubes,''
JHEP {\bf 0107}, 011 (2001)
[arXiv:hep-th/0106012].
}

\lref\BenaWP{ I.~Bena, ``The polarization of F1 strings into D2
branes: 'Aut Caesar aut  nihil','' Phys.\ Rev.\ D {\bf 67}, 026004
(2003) [arXiv:hep-th/0111156].
}

\lref\MateosPR{
D.~Mateos, S.~Ng and P.~K.~Townsend,
``Tachyons, supertubes and brane/anti-brane systems,''
JHEP {\bf 0203}, 016 (2002)
[arXiv:hep-th/0112054].
}

\lref\Buscher{T.~H.~Buscher, Phys.\ Lett.\ B {\bf 159}, 127
(1985),\ B {\bf 194}, 59 (1987),\ B  {\bf 201}, 466 (1988),
}

\lref\MeessenQM{ P.~Meessen and T.~Ortin, ``An Sl(2,Z) multiplet
of nine-dimensional type II supergravity theories,'' Nucl.\ Phys.\
B {\bf 541}, 195 (1999) [arXiv:hep-th/9806120].
}

\lref\don{
G.~T.~Horowitz and D.~Marolf,
``Counting states of black strings with traveling waves. II,''
Phys.\ Rev.\ D {\bf 55}, 846 (1997)
[arXiv:hep-th/9606113].
}

\lref\aspects{
M.~Kruczenski, R.~C.~Myers, A.~W.~Peet and D.~J.~Winters,
``Aspects of supertubes,''
JHEP {\bf 0205}, 017 (2002)
[arXiv:hep-th/0204103].
}

\lref\CallanHN{ C.~G.~Callan, J.~M.~Maldacena and A.~W.~Peet,
``Extremal Black Holes As Fundamental Strings,'' Nucl.\ Phys.\ B
{\bf 475}, 645 (1996) [arXiv:hep-th/9510134].
}

\lref\DabholkarNC{ A.~Dabholkar, J.~P.~Gauntlett, J.~A.~Harvey and
D.~Waldram, ``Strings as Solitons \& Black Holes as Strings,''
Nucl.\ Phys.\ B {\bf 474}, 85 (1996) [arXiv:hep-th/9511053].
}

\lref\CveticXH{ M.~Cvetic and F.~Larsen, ``Near horizon geometry
of  rotating black holes in five dimensions,'' Nucl.\ Phys.\ B
{\bf 531}, 239 (1998) [arXiv:hep-th/9805097].
}

\lref\EmparanWN{
R.~Emparan and H.~S.~Reall,
``A rotating black ring in five dimensions,''
Phys.\ Rev.\ Lett.\  {\bf 88}, 101101 (2002)
[arXiv:hep-th/0110260].
}

\lref\ReallBH{
H.~S.~Reall,
``Higher dimensional black holes and supersymmetry,''
Phys.\ Rev.\ D {\bf 68}, 024024 (2003)
[arXiv:hep-th/0211290].
}

\lref\ElvangMJ{
H.~Elvang and R.~Emparan,
 ``Black rings, supertubes, and a stringy resolution of black hole
non-uniqueness,''
JHEP {\bf 0311}, 035 (2003)
[arXiv:hep-th/0310008].
}

\lref\EmparanWY{
R.~Emparan,
``Rotating circular strings, and infinite non-uniqueness of black rings,''
arXiv:hep-th/0402149.
}

\lref\BenaWV{ I.~Bena, ``Splitting hairs of the three charge black
hole,'' arXiv:hep-th/0404073.
}

\lref\BenaWT{ I.~Bena and P.~Kraus, ``Three charge supertubes and
black hole hair,'' arXiv:hep-th/0402144.
}

\lref\ElvangRT{ H.~Elvang, R.~Emparan, D.~Mateos and H.~S.~Reall,
``A supersymmetric black ring,'' arXiv:hep-th/0407065.
}

\lref\BalasubramanianRT{ V.~Balasubramanian, J.~de Boer,
E.~Keski-Vakkuri  and S.~F.~Ross, ``Supersymmetric conical
defects: Towards a string theoretic description  of
black hole formation,''
Phys.\ Rev.\ D {\bf 64}, 064011 (2001) [arXiv:hep-th/0011217].
}

\lref\MaldacenaDR{ J.~M.~Maldacena and L.~Maoz,
``De-singularization  by rotation,'' JHEP {\bf 0212}, 055 (2002)
[arXiv:hep-th/0012025].
}

\lref\BW{ I.~Bena and N.~P.~Warner, ``One Ring to Rule Them All
... and  in the Darkness Bind Them?,'' arXiv:hep-th/0408106.
}

\lref\BenaWV{ I.~Bena, ``Splitting hairs of the three charge black
hole,'' arXiv:hep-th/0404073.
}

\lref\EEMRa{ H.~Elvang, R.~Emparan, D.~Mateos and H.~S.~Reall,
``A supersymmetric black ring,'' arXiv:hep-th/0407065.
}

\lref\EEMRb{ H.~Elvang, R.~Emparan, D.~Mateos and H.~S.~Reall,
``Supersymmetric black rings and three-charge supertubes,''
arXiv:hep-th/0408120.
}

\lref\BenaWT{ I.~Bena and P.~Kraus, ``Three charge supertubes and
black hole hair,'' arXiv:hep-th/0402144.
}

\lref\GauntlettWH{ J.~P.~Gauntlett and J.~B.~Gutowski,
``Concentric  black rings,'' arXiv:hep-th/0408010.
}

\lref\GauntlettQY{ J.~P.~Gauntlett and J.~B.~Gutowski, ``General
Concentric  Black Rings,'' arXiv:hep-th/0408122.
}

\lref\ElvangYY{ H.~Elvang, ``A charged rotating black ring,''
Phys.\ Rev. \ D {\bf 68}, 124016 (2003) [arXiv:hep-th/0305247].
}

\lref\StromingerSH{ A.~Strominger and C.~Vafa, ``Microscopic
Origin  of the Bekenstein-Hawking Entropy,'' Phys.\ Lett.\ B {\bf
379}, 99 (1996) [arXiv:hep-th/9601029].
}

\lref\ConstableDJ{ N.~R.~Constable, C.~V.~Johnson and R.~C.~Myers,
``Fractional branes and the entropy of 4D black holes,'' JHEP {\bf
0009}, 039 (2000) [arXiv:hep-th/0008226].
}

\lref\JohnsonGA{ C.~V.~Johnson, R.~R.~Khuri and R.~C.~Myers,
``Entropy of 4D Extremal Black Holes,''
Phys.\ Lett.\ B {\bf 378}, 78 (1996) [arXiv:hep-th/9603061].
}

\lref\MaldacenaGB{ J.~M.~Maldacena and A.~Strominger,
``Statistical  Entropy of Four-Dimensional Extremal Black Holes,''
Phys.\ Rev.\ Lett.\  {\bf 77}, 428 (1996) [arXiv:hep-th/9603060].
}

\lref\CveticXH{ M.~Cvetic and F.~Larsen, ``Near horizon geometry
of  rotating black holes in five dimensions,'' Nucl.\ Phys.\ B
{\bf 531}, 239 (1998) [arXiv:hep-th/9805097].
}

\lref\GiustoID{ S.~Giusto, S.~D.~Mathur and A.~Saxena, ``Dual
geometries for a set of 3-charge microstates,''
arXiv:hep-th/0405017.
}

\lref\GiustoIP{ S.~Giusto, S.~D.~Mathur and A.~Saxena, ``3-charge
geometries and their CFT duals,'' arXiv:hep-th/0406103.
}

\lref\LarsenUK{ F.~Larsen and E.~J.~Martinec, ``U(1) charges and
moduli in the D1-D5 system,'' JHEP {\bf 9906}, 019 (1999)
[arXiv:hep-th/9905064].
}

\lref\SeibergXZ{ N.~Seiberg and E.~Witten, ``The D1/D5 system and
singular CFT,'' JHEP {\bf 9904}, 017 (1999)
[arXiv:hep-th/9903224].
}

\lref\LuninUU{ O.~Lunin, ``Adding momentum to D1-D5 system,'' JHEP
{\bf 0404}, 054 (2004) [arXiv:hep-th/0404006].
}

\lref\KalloshUY{ R.~Kallosh and B.~Kol, ``E(7) Symmetric Area of
the Black Hole Horizon,'' Phys.\ Rev.\ D {\bf 53}, 5344 (1996)
[arXiv:hep-th/9602014].
}

\lref\BertoliniYA{ M.~Bertolini and M.~Trigiante, ``Microscopic
entropy of the most general four-dimensional BPS black  hole,''
JHEP {\bf 0010}, 002 (2000) [arXiv:hep-th/0008201].
}
\lref\BertoliniEI{ M.~Bertolini and M.~Trigiante, ``Regular BPS
black  holes: Macroscopic and microscopic description of the
generating solution,''
Nucl.\ Phys.\ B {\bf 582}, 393 (2000) [arXiv:hep-th/0002191].
}
\lref\MaldacenaDE{ J.~M.~Maldacena, A.~Strominger and E.~Witten,
``Black hole entropy in M-theory,'' JHEP {\bf 9712}, 002 (1997)
[arXiv:hep-th/9711053].
}

\lref\DavidWN{ J.~R.~David, G.~Mandal and S.~R.~Wadia,
``Microscopic  formulation of black holes in string theory,''
Phys.\ Rept.\ {\bf 369}, 549 (2002) [arXiv:hep-th/0203048].
}

\lref\AharonyTI{ O.~Aharony, S.~S.~Gubser, J.~M.~Maldacena,
H.~Ooguri and  Y.~Oz, ``Large N field theories, string theory and
gravity,'' Phys.\ Rept.\  {\bf 323}, 183 (2000)
[arXiv:hep-th/9905111].
}

\lref\BanadosWN{ M.~Banados, C.~Teitelboim and J.~Zanelli, ``The
Black hole  in three-dimensional space-time,'' Phys.\ Rev.\ Lett.\
{\bf 69}, 1849 (1992) [arXiv:hep-th/9204099].
}

\lref\BrownNW{ J.~D.~Brown and M.~Henneaux, ``Central Charges In
The  Canonical Realization Of Asymptotic Symmetries: An
Example From Three-Dimensional Gravity,''
Commun.\ Math.\ Phys.\  {\bf 104}, 207 (1986).
}

\lref\KutasovZH{ D.~Kutasov, F.~Larsen and R.~G.~Leigh, ``String
theory in magnetic  monopole backgrounds,'' Nucl.\ Phys.\ B {\bf
550}, 183 (1999) [arXiv:hep-th/9812027].
}

\lref\LarsenDH{ F.~Larsen and E.~J.~Martinec, ``Currents and
moduli in the (4,0)  theory,'' JHEP {\bf 9911}, 002 (1999)
[arXiv:hep-th/9909088].
}

\lref\min{ J.~P.~Gauntlett, J.~B.~Gutowski, C.~M.~Hull,
S.~Pakis and H.~S.~Reall,
``All supersymmetric solutions of minimal supergravity in five dimensions,''
Class.\ Quant.\ Grav.\  {\bf 20}, 4587 (2003)
[arXiv:hep-th/0209114].

J.~B.~Gutowski and H.~S.~Reall, ``General
supersymmetric AdS(5) black holes,'' JHEP {\bf 0404}, 048 (2004)
[arXiv:hep-th/0401129].

J.~B.~Gutowski, D.~Martelli and H.~S.~Reall,
``All supersymmetric solutions of minimal supergravity in six
dimensions,'' Class.\ Quant.\ Grav.\  {\bf 20}, 5049 (2003)
[arXiv:hep-th/0306235].
}

\lref\usc{
C.~N.~Gowdigere, D.~Nemeschansky and N.~P.~Warner,
``Supersymmetric solutions with fluxes from algebraic Killing spinors,''
arXiv:hep-th/0306097.

K.~Pilch and N.~P.~Warner,
``Generalizing the N = 2 supersymmetric RG flow solution of IIB
supergravity,''
Nucl.\ Phys.\ B {\bf 675}, 99 (2003)
[arXiv:hep-th/0306098].
}

\lref\KS{
I.~R.~Klebanov and M.~J.~Strassler,
``Supergravity and a confining gauge theory: Duality cascades and
$\chi$SB-resolution of naked singularities,''
JHEP {\bf 0008}, 052 (2000)
[arXiv:hep-th/0007191].

I.~R.~Klebanov and A.~A.~Tseytlin,
``Gravity duals of supersymmetric SU(N) x SU(N+M) gauge theories,''
Nucl.\ Phys.\ B {\bf 578}, 123 (2000)
[arXiv:hep-th/0002159].
}

\Title{
  \vbox{\baselineskip12pt \hbox{hep-th/0408186}
  \hbox{UCLA-04-TEP-35}
  \vskip-.5in}
}{\vbox{
  \centerline{Microscopic Description of Black Rings in AdS/CFT}
 }}

\centerline{Iosif Bena and Per Kraus}

\bigskip\medskip
\centerline{ \it Department of Physics and Astronomy,
UCLA, Los Angeles, CA 90095-1547,
USA}

\medskip
\medskip
\medskip
\medskip
\medskip
\medskip
\baselineskip14pt
\noindent

We discuss some aspects of the recently discovered BPS black ring
solutions in terms of the AdS/CFT correspondence.  In the type IIB
frame in which the black ring carries the charges of the D1-D5-P
system, we propose a microscopic description of the rings in the
orbifold CFT governing this system. In our proposal, the CFT
effectively splits into two parts: one part captures the
supertube-like properties of the ring, and the other captures the
entropy. We can also understand the black ring entropy by relating
the geometry near the ring to BPS black holes in four dimensions,
although this latter approach  does not directly lead to an
identification of black rings in terms of the D1-D5-P CFT.

\Date{August, 2004}
\baselineskip14pt
\newsec{Introduction}

An interesting new solution of five dimensional minimal
supergravity was  discovered recently \EEMRa: a supersymmetric
black ring with horizon topology $S^1 \times S^2$.  More general
solutions in eleven dimensional supergravity were subsequently
obtained in \refs{\BW,\EEMRb}, as well as multi-ring solutions in
\refs{\GauntlettWH,\GauntlettQY}.  These solutions have their
origins in prior studies of (non-BPS or singular) black rings
\refs{\EmparanWN,\ElvangYY,\ElvangMJ,\EmparanWY}, in studies of
three-charge supertubes \refs{\BenaWT,\BenaWV}, as well as in
methods for classifying supersymmetric solutions
\refs{\usc,\min}.

One of the most interesting features of these black rings is that,
even  after fixing all their conserved charges, they retain
several discrete parameters which can be varied; furthermore, the
area of the horizon depends on these parameters.  Therefore, these
black rings have ``hair''.  This makes it especially interesting
to see whether the entropy can be computed microscopically in
string theory in terms of the AdS/CFT correspondence, and to see
how the hair parameters are accounted for.  In previous
microscopic computations of black hole entropy following
\StromingerSH\ one counts up all states with given values of
conserved charges, but for the black rings one  apparently needs
to impose additional restrictions on the class of states to be
counted.

Mathur and collaborators
\refs{\LuninJY,\mathurstretch,\LuninBJ,\mathur,\LuninUU,\GiustoIP,\GiustoID}
have initiated a program
to argue that every microstate of the D1-D5-P system corresponds
to a particular bulk solution (though not necessarily a smooth and
classical solution of supergravity) with no event horizon. In this
picture, the logarithm of the black hole entropy is equal to the
total number of choices for the hair parameters. For some choices
of parameters, the black ring solutions have vanishing entropy,
and so can potentially serve as geometries dual to individual
microstates in the manner envisaged by Mathur. One of our goals
here is to identify these microstates.  We will also see
that despite having small curvature these supergravity solutions
are singular, as they have cycles that shrink to zero size; whether or not
the singularities can be resolved remains to be seen.

The black ring can be thought of as a  supertube
\refs{\supertube,\EmparanUX},  with three conserved charges and
three dipole charges \BenaWT. The dipole charges correspond to
branes with one direction describing a topologically trivial
closed curve in spacetime, and hence giving rise to no net charge.
In the type IIB duality frame in which the black rings carries the
conserved charges of the D1-D5-P system, the dipole charges are
those of D1-branes, D5-branes, and Kaluza-Klein monopoles.  Very
near the ring the dipole branes effectively appear flat.  Since
the dipole branes are precisely those appearing in the
construction of the four dimensional black hole
\refs{\MaldacenaGB,\JohnsonGA}, it comes as no surprise that the
black ring entropy is just a rewriting of the four dimensional
black hole entropy, as we discuss in the next section.

On the other hand, after taking the near horizon decoupling limit,
the black ring solutions are asymptotically AdS$_3 \times S^3
\times T^4$, and so should be describable in terms of the usual
D1-D5 CFT.  Most of what we know about this CFT comes from working
at the point in moduli space in which it is free --- the
``orbifold point'' (see, e.g., \refs{\AharonyTI,\DavidWN}  for
reviews of the D1-D5 system and  \refs{\SeibergXZ,\LarsenUK} for
more discussion of the moduli space.)  Since the supergravity
approximation is invalid at the orbifold point, direct comparisons
between the two sides of the duality can only be made in certain
cases.
As we will see, the black ring solutions
can be thought of as an amalgam
of the two-charge supertube and the BMPV \bmpv\ black hole, and
since we know that many of their properties are correctly captured
at the orbifold point we can hope that the success carries over.

In particular, two-charge supertubes are described in terms of
multiple ``component strings'' with angular momentum provided by
fermion zero modes \refs{\LuninJY,\LuninIZ}, while the BMPV black
hole is represented as a single component string with the angular
momentum carried by fermion momentum modes \bmpv. With this in
mind, we propose  a microscopic description for the black rings in
which the effective string of the D1-D5 system effectively splits
into two parts, one corresponding to the supertube and one to the
BMPV black hole.  After making one phenomenological assumption
about the length of the component strings, we are able to explain
the entropy of all circular black rings. It will be interesting to
see whether this assumption can eventually be derived from first
principles. 

Another way to view  the microscopic description of the
black rings is to regard the supergravity solutions as describing
an RG flow. At the AdS$_3$ boundary one has the CFT of the D1-D5
system, which has $(4,4)$ supersymmetry and $c_{_{\!  \, U \! V}} = 6 N_{D1}
N_{D5}$, while near the ring one has the CFT of the
four-dimensional black hole, which is a $(4,0)$ theory with
$c_{_{\! \, I \! R}} =6n_{d1} n_{d5} n_{kk}$ (lower case letters denote the
dipole charges).  We will analyze the geometry at the IR end of
the flow near the ring, but understanding the full flow from the
boundary gauge theory point of view is left as an open question
for the future.

Some of the observations we make below can also be found in
\EEMRb,  which appeared while this paper was being written.

\newsec{The IIB Solution}

The supersymmetric black ring was given an M-theory description in
\BW\EEMRb\ in  terms of intersecting M2-branes and M5-branes.
For AdS/CFT purposes it is more convenient to work in the duality
frame in which the solution carries the same charges as the
familiar D1-D5-P system. In this frame, the solution will also
carry dipole ``charges'' of D1-branes, D5-branes, and KK-monopoles
(this IIB solution also appears in \EEMRb, and its geometry was
thoroughly investigated there)

To reach  the
D1-D5-P  frame we compactify the M-theory solution
along one of the M2 branes, and T-dualize three more times.
We obtain a solution of type IIB supergravity compactified on $T^5$, where
$x^{6,7,8,9}$ describe a $T^4$ of volume $V_4$, and $x^5$ is a
circle of radius $R_{KK}$.  The number of integral  units of
charge and their orientations are
\eqn\aa{ N_1 ~~ D1(5), \quad N_2 ~~ D5(56789), \quad N_3 ~~
P(5)~.}
The dipole branes have one common worldvolume direction
curled up into a circle, which we will now denote as $x$ (this is
the ring direction of the black ring).  The number and orientation
of the dipole branes is then
\eqn\ab{ n_1~~ d5(x6789), \quad n_2~~ d1(x), \quad n_3~~
kk(x56789)}
where we are denoting dipole quantities by lower case letters.
$x^5$ is the special KK circle of the kk-monopole.

The quantized charges above are related to the parameters
appearing in the metric as
\eqn\ac{\eqalign{  Q_1 &= {(2 \pi)^4 g \ap^3  \over V_4}N_1, \quad
Q_2 =g \ap N_2, \quad  Q_3 = {(2\pi)^4 g^2 \ap^4 \over V_4
R_{KK}^2}N_3 , \cr q_1 & ={g \ap \over R_{KK} }n_1, \quad  q_2
={(2\pi)^4 g\ap^3 \over V_4 R_{KK}}n_2, \quad  q_3  = R_{KK}n_{3}.
}}

The charges $N_i$ are measured as flux integrals at infinity.
These are to be distinguished from the charges measured by flux
integrals at the ring itself.  The latter are not conserved in
general, but do tell us more directly what the ring is made of.
Denoting these by $\Nb_i$, they are given by (as will be seen from
the explicit form of the solution)
\eqn\ad{ \Nb_1 = N_1 - n_2n_3, \quad \Nb_2 = N_2 - n_1 n_3, \quad
\Nb_3 = N_3 - n_1 n_2.}
We also have the corresponding definitions $\Qb_1 =Q_1 - q_2 q_3$,
etc. As one can see from the equations governing the black ring solutions \BW,
the difference between the asymptotic and near-ring charges
comes from the charge carried by supergravity fields. This feature of the black
ring solution is very similar to the Klebanov-Strassler solution \KS, which also contains
D-brane charges ``dissolved'' into fluxes via the Chern-Simons terms in the action.

The metric in string frame is
\eqn\ae{ds^2 = -(e_1)^2+(e_5)^2 + \sqrt{Z_1 Z_2}ds_{4}^2 +
\sqrt{{Z_1 \over Z_2}}ds_{T^4}^2}
where
\eqn\af{\eqalign{e_1&= {1 \over Z_1^{1/4} Z_2^{1/4} Z_3^{1/2} }
(dt+k_\psi d\psi  + k_\phi d\phi) \cr e_5 & = -e_1 + {Z_3^{1/2}
\over Z_1^{1/4} Z_2^{1/4}} (dt+  dx_5 - s_\psi d\psi - s_\phi
d\phi)~. }}
The solution is easiest to express if we use a coordinate system \EEMRa\
in which the metric of flat $\IR^4$ is
\eqn\ag{\eqalign{ds_4^2 & = {R^2 \over (x-y)^2}\left[ {dy^2 \over
y^2-1}+(y^2-1)d\psi^2 +{dx^2 \over 1-x^2}+(1-x^2)d\phi^2\right]
\cr & = d\rho^2 + \rho^2(d\theta^2 + \sin^2\theta d\psi^2 + \cos^2
\theta d\phi^2) }}
related to the usual flat $\IR^4$ coordinates by
\eqn\ah{\eqalign{ \rho \sin \theta & = {\sqrt{y^2-1}\over x-y}R,
\quad\quad \rho \cos\theta = {\sqrt{1-x^2}\over x-y}R \cr x& =
-{\rho^2 -R^2 \over \Sigma}, \quad\quad y = -{\rho^2+R^2 \over
\Sigma} }}
with
\eqn\ai{\Sigma = \sqrt{(\rho^2-R^2)^2+4 R^2 \rho^2 \cos^2 \theta}~.}
The coordinate ranges are $-1 \leq x \leq 1$, $-\infty \leq y \leq -1$,
and $\psi$ and $\phi$ are $2 \pi$ periodic.

In these coordinates, the $Z_i$ functions (which would be harmonic
in the absence of the dipole charges) are
\eqn\aj{\eqalign{Z_1 &= 1 + {\Qb_1 \over 2R^2}(x-y) - {q_2 q_3
\over 4R^2}(x^2-y^2)\cr &= 1 + {\Qb_1 \over\Sigma} +{q_2 q_3
\rho^2 \over \Sigma^2} ~,}}
with $Z_{2,3}$ given by cyclic permutations of the labels
$(123)$.  The other quantities appearing in the solution are
\eqn\ak{\eqalign{ k_\psi & = -\half(q_1+q_2+q_3)(1+y) -{1 \over
8R^2}(1-y^2)\left[ \sum_i q_i \Qb_i - q_1q_2q_3(x+y)\right]\cr
&\quad = {2 (q_1+q_2+q_3)R^2 \rho^2 \sin^2\theta \over \Sigma
(\rho^2+R^2 +\Sigma)}+{\rho^2 \sin^2 \theta \over 2
\Sigma^2}\left[\sum_i q_i \Qb_i +2q_1q_2q_3 {\rho^2 \over
\Sigma}\right] \cr k_\phi & = -{1 \over 8R^2}(1-x^2)\left[ \sum_i
q_i \Qb_i - q_1q_2q_3(x+y)\right]= -{\rho^2 \cos^2 \theta \over 2
\Sigma^2}\left[\sum_i q_i \Qb_i +2q_1q_2q_3 {\rho^2 \over
\Sigma}\right]\cr   s_\psi&= \half q_3 (1+y)= \half q_3 {\Sigma
-\rho^2 -R^2 \over \Sigma} \cr s_\phi &= -\half q_3 (1+x)=- \half
q_3 {\Sigma -\rho^2 +R^2 \over \Sigma}.}}
The dilaton is
\eqn\al{e^{-2\Phi} = {Z_2 \over Z_1}.}
The solution carries the angular momenta
\eqn\am{\eqalign{J_\phi & =-{R_{KK} V_4 \over 2(2 \pi)^4 \ap^4
g^2}\left( \sum_i q_i \Qb_i +2 q_1 q_2 q_3 \right)=-{1 \over
2}\sum_i n_i \Nb_i -n_1n_2n_3 \cr J_\psi & =-J_\phi+ {R_{KK} V_4
\over (2 \pi)^4 \ap^4 g^2}\left( q_1+q_2+ q_3\right)R^2~.
 }}

There is a black hole horizon located at $y=-\infty$, and the
corresponding black hole entropy works out to be
\eqn\an{\eqalign{ S = {A \over 4 G}  & = 2\pi \Big[ -{1 \over
4}(n_1^2 \Nb_1^2 +n_2^2 \Nb_2^2+n_3^2 \Nb_3^2)+{1 \over 2}(n_1n_2
\Nb_1 \Nb_2 + n_1 n_3 \Nb_1 \Nb_3 + n_2 n_3 \Nb_2 \Nb_3) \cr &
\quad\quad\quad- n_1 n_2 n_3 (J_\psi +J_\phi)\Big]^{1/2}. }}
One of our main goals is to make as much progress as we can in
understanding the formula \an\ from the point of view of the
AdS/CFT correspondence.  In the near horizon limit which we take
below, the ring solution is seen to be asymptotically $AdS_3
\times S^3 \times T^4$, and so should be describable as a state
(or ensemble of states) in the same boundary theory as the usual
D1-D5-P system.

There is actually another, simpler, way to understand the entropy
formula. Instead of taking the dipole branes to be curled into a
circle we can take them to be flat, as was done in \BenaWV. The
resulting brane configuration has $6$ spatial worldvolume
directions, and if we wrap these on $T^6$ then we obtain a four
dimensional black hole preserving $4$ supercharges.  The general
entropy for this class of black holes is \KalloshUY
\eqn\ana{ S = 2\pi \sqrt{J_4}~,}
where $J_4$ is the quartic $E_{7(7)}$ invariant, which can be
expressed in the basis $(x_{ij},y^{ij})$ as
\eqn\ao{\eqalign{ J_4 &= -{1 \over 4}(x_{12}y^{12} + x_{34}y^{34}
+x_{56}y^{56}+x_{78}y^{78})^2-
(x_{12}x_{34}x_{56}x_{78}+y^{12}y^{34}y^{56}y^{78})\cr & +
x_{12}x_{34}y^{12}y^{34}+ x_{12}x_{56}y^{12}y^{56} +
x_{34}x_{56}y^{34}y^{56}+x_{12}x_{78}y^{12}y^{78}+ x_{34}x_{78}
y^{34}y^{78}\cr &+x_{56}x_{78}y^{56}y^{78} .}}
We note that \ana\ agrees with \an\ under the identifications
\eqn\app{\eqalign{x_{12}&=\Nb_1, \quad x_{34}=\Nb_2, \quad
x_{56}=\Nb_3,\ \quad x_{78}=0, \cr  y^{12} &= n_1, \quad
y^{34}=n_2, \quad y^{56} = n_3, \quad y^{78}=J_\psi +J_\phi.}}
These identifications can be read off by starting from the
M-theory version of the near-ring solution as presented in
\refs{\BenaWV,\BW,\EEMRb}, reducing to IIA along the momentum
direction, and then comparing with Table 2 of \BertoliniEI. We
should also notice that $x_{78}$, which is zero in our case,
corresponds to having a KK monopole whose special direction is
along the ring. It will be interesting to see if the black ring
solutions can be generalized to include this charge.

 A microscopic
computation of the entropy was given in \refs{\BertoliniEI,
\BertoliniYA}  following the approach of \MaldacenaDE.  So in this
sense, the entropy formula \an\ is understood in terms of the
theory living on the dipole branes. However, our main goal is to
understand the entropy in terms of the D1-D5-P CFT, so that we can
have a common understanding of the black ring and the usual
D1-D5-P black hole.

\newsec{Decoupling Limit}

The black ring geometry has two physically interesting limits. One
is obtained exactly like in the three charge black hole case, by
removing the asymptotically flat region of the solution to obtain
a near-horizon geometry dual to the D1-D5-P CFT. The other limit
is obtained by zooming in further on the near-ring region. In this
limit the metric simply becomes the flat ring metric.

We leave the analysis of the near-ring limit to section 6, and focus for now
on the  near-horizon decoupling limit.  As usual, we
wish to take $\ap \rightarrow 0$ while scaling coordinates and
moduli such that the metric has an overall factor of $\ap$ (so
that the action $S \sim {1 \over G} \int \! \sqrt{-g} R$ is finite
in the limit).  In the $(x,y,\psi,\phi)$ coordinates this is
achieved by
\eqn\ba{ V_4 \sim \ap^2 ,\quad R \sim \ap}
with all other coordinates and moduli fixed.  From \ac\ this
implies the following scaling of the charges
\eqn\bb{Q_{1,2} \sim \ap ,\quad Q_3 \sim \ap^2, \quad q_{1,2} \sim
\ap, \quad q_3 \sim \ap^0~.}
Note that $Q_1 \sim q_2 q_3$ (and permutations thereof) and so
$\Qb_i$ scale the same as $Q_i$.  Inserting these scalings into
the $Z_i$, we see that we should drop the $1$ from $Z_{1,2}$ while
retaining it for $Z_3$; this is the same as in the standard
D1-D5-P case.  Finally, since $q_{1,2}$ scale to zero compared to
$q_3$, whenever we see the combination $(q_1 + q_2 +q_3)$, as in
\ak\ and \am, we should replace it by $q_3$.

Upon taking the near-horizon limit, the large $\rho$ behavior of
the metric is
\eqn\bc{\eqalign{ ds^2 &\approx {\rho^2 \over \sqrt{Q_1
Q_2}}(-dt^2 +dx_5^2) + \sqrt{Q_1 Q_2} {1 \over \rho^2} d\rho^2 +
\sqrt{Q_1 Q_2}(d\theta^2 + \sin^2 \theta d\psi^2 + \cos^2 \theta
d\phi^2) \cr &+ \sqrt{{Q_1 \over Q_2}} ds_{T^4}^2}}
which is the usual AdS$_3 \times S^3 \times T^4$ with $\ell_{AdS}
= (Q_1 Q_2)^{1/4}$.   Therefore, the black ring should be
describable in terms of the CFT of the D1-D5-P system.

 We recall
that the angular momentum on the $S^3$ becomes the $SO(4) \approx
SU(2)_L \times SU(2)_R$ R-charge symmetry of the CFT.  Left and
right moving fermions transform as doublets under the
corresponding $SU(2)$ factor.   The diagonal $SU(2)$ generators,
normalized to have integer eigenvalues,  are related to
$J_{\psi,\phi}$ by
\eqn\bd{J_L = J_\psi -J_\phi, \quad J_R = J_\psi+ J_\phi.}

Henceforth, our discussion of the supergravity solutions will be
strictly in the context of the near horizon limit.

\newsec{Special cases}

In order to identify the black ring in the CFT it is helpful to
first review some simple, well understood, special cases.

\subsec{ D1-D5 $\rightarrow$ kk supertube}

Set $Q_3 =q_1 =q_2 =0$, leaving us with $N_1$ D1-branes, $N_2$
D5-branes and $n_3$ kk monopoles, the latter being a dipole
charge. These solutions have been intensively studied
\refs{\BalasubramanianRT,\MaldacenaDR,\emparan,\LuninJY,\LuninIZ}.
In order to avoid singularities
 or closed timelike curves  we must set the ring
radius to be
\eqn\ca{ R = {\sqrt{Q_1 Q_2} \over q_3}~,}
which gives
\eqn\cb{J_L =J_R = {N_1 N_2 \over n_3}~.}
These solutions are interpreted as being particular ground states
in the Ramond sector of the CFT.

In particular, at the orbifold point in the  moduli space, we can
think of the CFT as being a string of length $2\pi N_1 N_2$, which
can be broken up into ``component strings'', each of length $2\pi$
times an integer.  On each component string live $4$ real bosons,
and $4$ real right-moving and left-moving fermions.  Two of the
right-moving fermions carry $J_R = +1$, while the other two carry
$J_R =-1$, and analogously for the left-movers.  Upon quantizing
the zero mode fermions we find that each component string carries
integer charges with $|J_{L,R}| \leq 1$.

The state \cb\ corresponds to taking all component strings  to
have  length $2\pi n_3$ and carrying $J_L = J_R =1$, so that there
are a total of $N_1 N_2/ n_3$ components in total, carrying the net
charge \cb. The interpretation of more general states is spelled
out in detail in \LuninJY\MaldacenaDR.

\subsec{BMPV black hole}

Setting $R=0$ gives the BMPV black hole \bmpv, which carries
\eqn\cc{ J_L = \sum_i n_i \Nb_i +2n_1 n_2 n_3, \quad J_R =0~.}
The entropy of the black hole is
\eqn\cd{ S = 2\pi \sqrt{N_1 N_2 N_3 - J_L^2/4}.}
Note that this cannot be obtained directly from the entropy
formula \an,  since the $R\rightarrow 0$ limit is discontinuous in
the geometry. Upon defining the new angular coordinates \CveticXH
\eqn\cda{\eqalign{\psit& = \psi - {J_L \over 2N_1 N_2
R_{KK}}(t+x_5) \cr\phit& = \phi + {J_L \over 2N_1 N_2
R_{KK}}(t+x_5)}}
the metric becomes
\eqn\ce{\eqalign{ ds^2 &= {\rho^2 \over \sqrt{Q_1 Q_2}}(-dt^2
+dx_5^2) + \sqrt{Q_1 Q_2} {1 \over \rho^2} d\rho^2 +{Q_1 Q_2 Q_3
-C^2 J_L^2/4 \over (Q_1 Q_2)^{3/2}}(dt+dx_5)^2 \cr &~~+\sqrt{Q_1
Q_2}(d\theta^2 + \sin^2 \theta d\psit^2 + \cos^2 \theta d\phit^2)
+ \sqrt{{Q_1 \over Q_2}} ds_{T^4}^2}}
with
\eqn\cf{ C^2 =\left({(2\pi)^4 g^2 \ap^4 \over R_{KK}V_4}\right)^2=
{Q_1 Q_2 Q_3 \over N_1 N_2 N_3 }.}
The $(\rho,t,x_5)$ part of the metric is that of an extremal,
rotating  BTZ black hole \BanadosWN.

The entropy \cd\ is most readily obtained as follows.  We have a
${\cal N}=(4,4)$ SCFT of central charge $c=6N_1 N_2$ and we wish
to count the number of left-moving states in the Ramond sector
with $L_0 = N_3$ and with the R-charge $J_L$.  We can apply a
spectral flow transformation to remove the R-charge.  A spectral
flow by $\eta_L$ acts as
\eqn\cg{\eqalign{ L_0 &\rightarrow L_0 +\eta_L J_L +{c\over
6}\eta_L^2 \cr J_L &\rightarrow J_L +{c\over 3}\eta_L~.}}
If we take $\eta_L = -{J_L \over 2 N_1 N_2}$ then we get $J_L=0$
and $L_0 =N_3 - {J_L^2 \over 4 N_1 N_2}. $  Cardy's formula, $S=
2\pi \sqrt{{c \over 6} L_0}$ then yields \cd.

The coordinate transformation \cd\ is the bulk implementation of the
spectral flow transformation \BalasubramanianRT\MaldacenaDR, as is
evident from the transformed level appearing in the BTZ metric.
We can also read off another important feature. In the new
coordinates the periodicity $(x_5,\psi,\phi) \cong (x_5 + 2\pi
R_{KK},\psi,\phi)$  turns into
\eqn\ch{(x_5,\psit,\phit) ~\cong ~(x_5 +2 \pi R_{KK}, \psit- {J_L
\over 2 N_1 N_2}2\pi,\phit+ {J_L \over 2 N_1 N_2}2\pi).}
Since ${J_L \over N_1 N_2}$ is not in general an integer, the
effective length of the $x_5$ direction (i.e. the length appearing
in the quantization condition for momenta in this direction) is
$N_1 N_2$ times its naive value\foot{The factor of $\half$ which
we ignored is due to the anti-periodicity of fermions.}. So in
terms of the component string picture of the (orbifold point) of
the CFT, the BMPV black hole corresponds to having a single
component string of length $2\pi N_1 N_2$.  Since there is only a
single component string, the R-charge (or equivalently, the
angular momentum on $S^3$) is carried by the momentum modes of the
fermions, rather than just by the zero modes as for the supertube.

\newsec{Black ring entropy}

Examination of the structure of the general solution indicates
that it combines features of the supertube and the BMPV black
hole, and so it is natural to seek a CFT description which
similarly combines their microscopic elements.  To this end, we
first separate the angular momenta into two parts
\eqn\da{\eqalign{{\rm Tube}:\quad & J_\T = J_\psi+J_\phi = {q_3
R^2 \over C}\cr {\rm BMPV:}\quad &J_\B = J_\phi = -\half \sum_i
n_i \Nb_i -n_1 n_2 n_3~,}}
where $C$ was defined in \cf.  Our basic proposal is that the
effective string of length $2\pi N_1 N_2$  splits into
two parts, which we call the tube string and the BMPV string, with
lengths $2\pi L_\T$ and $2\pi L_\B$.  The tube string is broken up
into a number of component strings of equal length $2 \pi \ell_c$, and
carries $J_\T$ in the fermion zero modes.  The BMPV string, on the
other hand, consists of a single component and carries the
momentum, entropy, and $J_\B$ as momentum excitations.

Since each component of the tube string carries $J_\T = 1$, we
immediately find that the number of such components is
\eqn\db{ {L_\T \over \ell_c} = J_\T.}
Since we also have $L_\T + L_\B =  N_1 N_2$, there is only one
free parameter remaining, which we can take to be $\ell_c$.  A
complete explanation would include a computation of $\ell_c$ from
first principles; we will not be able to achieve this in general,
but we will see how to understand the form of $\ell_c$ in limiting
cases.

Given our interpretation, the entropy should take the BMPV form
\eqn\dc{S =2\pi \sqrt{L_\B N_3 - J_\B^2}~.}
Next, in order to put the black ring entropy in a more suggestive
form,  we parameterize the angular momenta as
\eqn\dd{J_\T = {\Nb_1 \Nb_2 \over n_3} -\delta, \quad J_\B =-n_3
N_3 +\gamma~,}
 which, after some algebra, brings the entropy \an\ to
the form
\eqn\de{S = 2\pi \sqrt{ n_1 n_2 n_3 \delta - \gamma^2}~.}
Another useful formula for $\gamma$ is
\eqn\dea{\gamma= \half(n_3 \Nb_3 -n_1 \Nb_1 -n_2 \Nb_2)~.}

We now examine some particular cases of increasing complexity.

\subsec{Case 1:~ $\delta = \gamma=0$}

These are zero entropy solutions, and so should correspond to
individual CFT microstates.  In fact, these states are nicely
understood by taking the length of the component strings to be
\eqn\df{\ell_c = n_3}
which is the same length as for the pure supertube, as discussed
in section 4.  From \db\ we have $L_\T = \Nb_1 \Nb_2$, which
implies
\eqn\dg{L_\B = N_1 N_2 - \Nb_1 \Nb_2 =n_3^2 N_3~.}
We need to include fermionic excitations on the BMPV string in
order to account for $J_\B$.  In order to get the maximally
negative  $J$ we can fill up the Fermi sea with the negatively
charged fermions.  After a short calculation, one finds that the
total charge one obtains in this way is\foot{This is just the
usual bound on the BMPV angular momentum, as can be equivalently
obtained, for example, from the spectral flow transformation
discussed earlier.}
\eqn\dh{ J_\phi^{\rm max} = - \sqrt{L_\B N_3}.}
Given \dg, we see that $J_\phi^{\rm max} =-n_3 N_3= J_\B$, and so
the angular momentum is correctly accounted for.

Thus, these microstates nicely match up with the gravity
solutions. Furthermore, they provide examples of geometries dual
to microstates with nonzero D1, D5, and momentum charges.
However, as we'll see in the next section, these geometries fail
to cap off smoothly in the way that one needs in order to get a
completely smooth supergravity description.

\subsec{Case 2: $\delta \neq 0,~ \gamma=0$}

In this case the black ring has the nonzero entropy $S= 2\pi
\sqrt{n_1 n_2 n_3 \delta}$.   A first guess for how this comes
about is as follows.  To reduce $J_\T$ we can start converting the
tube component strings into the BMPV string.  If each component
string has length $\ell_c = n_3$ as above, then converting
$\delta$ such components increases $L_\B$ by $n_3 \delta$.  But
from \dc\ this gives a microscopic entropy $S_{{\rm micro}} = 2\pi
\sqrt{n_3 N_3 \delta} = 2\pi \sqrt{n_1 n_2 n_3 \delta + n_3 \Nb_3
\delta}$.  We see that this is too large, although it does give
the correct result in the regime $n_i \gg \Nb_i$.

In fact, there was no particularly good reason to assume that
$\ell_c$ remains unchanged for nonzero $\delta$.  We now simply
demand that $\ell_c$ change in such a way as to reproduce the
correct entropy.  Although this doesn't do much to explain the
entropy in this case, it does have the virtue that the same
formula for $\ell_c$ will continue to reproduce the entropy for
nonzero $\gamma$, which is a nontrivial result.

In particular, if we again assume there are $J_\T$ component tube
strings, each of length $\ell_c$, and that $L_\B = N_1 N_2 - J_\T
\ell_c$, then we obtain the correct entropy by taking
\eqn\di{ \ell_c = \left(1+ {\Nb_3 \over N_3 } {\delta \over
J_\T}\right)n_3~.}
One can in principle hope to test this identification by
performing the  sort of scattering experiments in \LuninJY,
but for now we leave it as a phenomenological assumption.

\subsec{Case 3: $\delta \neq 0,~ \gamma \neq 0$}

Proceeding as in Case $2$  but now for arbitrary $\gamma$, and in
particular using the {\it same} formula \di\ for the component
string lengths,  we can work out the combination $L_\B N_3 -
J^2_\B$ to find a nontrivial cancellation between terms linear in
$\gamma$. Hence the two entropies \dc\ and \de\ precisely match.
Thus, the phenomenological assumption \di\ correctly explains the
entropy of all circular black rings.

We have now identified some additional zero entropy microstates
corresponding to $\gamma^2 = n_1 n_2 n_3 \delta$.  Again, they
have zero entropy because the angular momentum bound is saturated,
so the BMPV string is described by a filled Fermi sea. It is
interesting to note that our supergravity solutions do not seem to
be able to capture zero entropy microstates whose component string
length is different from the value in \di. It would be nice to
understand why it is so.

\newsec{Near ring geometry}

We now examine the geometry in the region near the ring in each of
the three cases above.  The main point is that the solutions
approach AdS$_3 \times S^3/Z_{n_3} \times T^4$ near the ring,
which is the same geometry as for a collection of D1-branes,
D5-branes, and KK-monopoles; in other words, near the ring the
dipole branes dominate (as is clear from the form of the $Z_i$ for
small $\Sigma$).  Since one of the spatial Poincar\' e coordinates parallel
to the boundary of AdS$_3$ is compactified, the geometries will
all be singular whenever the area of the horizon shrinks to zero
size.  Thus, although we have been able to identify some
low curvature geometries dual to microstates of the D1-D5-P system, they are not
smooth supergravity solutions, since they do not cap off smoothly.
On the other hand, it is then clear that the problem has been
reduced to finding such smooth geometries for the $P=0$ limit of
the D1-D5-KK system; i.e. to finding smooth geometries
corresponding to the ground states of the $4D$ black hole.

\subsec{Case 1:~ $\delta = \gamma=0$}

If we write
\eqn\ea{\eqalign{ \psit & = \psi - {1 \over q_3} (x_5+t), \quad
\phit  = \phi +{1 \over q_3}(x_5+t) \cr \xt_5 & = q_3 \psi -t,
\quad \cos (2\alpha)=x, \quad \yt  = -\sqrt{{q_1 q_2 q_3^2 \over 2
Q_1 Q_2}} \sqrt{y}~, }}
then the leading form of the metric for $\yt\rightarrow \infty$ is
\eqn\eb{ds^2 = {\sqrt{q_1 q_2 q_3^2} \over \yt^2}(d\yt^2 -dt^2
+d\xt_5^2)+\sqrt{q_1 q_2 q_3^2}(d\alpha^2 + \sin^2 \alpha d\psit^2
+ \cos^2 \alpha d\phit^2)+ \sqrt{{q_2 \over q_1}}ds_{T^4}^2~.}
Note that we now have the angular identifications $(\psit, \phit)
\cong (\psit - {2\pi \over n_3}, \phit + {2\pi \over n_3})$,
giving rise to $S^3 / Z_{n_3}$.  We also have $\xt_5 \cong \xt_5 +
2\pi n_3 R_{KK}$.  The geometry is singular at $\yt= \infty$ since
the size of the $\xt_5$ circle shrinks to zero.

 The spacetime central charge
obtained from the Brown-Henneaux formula \BrownNW\ is
\eqn\ec{ c = {3 \ell_{\rm AdS} \over 2 G_3} =  6 n_1 n_2 n_3~.}
To obtain this we took into account that $G_3$ is $n_3$ times
larger than it would be for AdS$_3 \times S^3$, due to the reduced
volume of $S^3 / Z_{n_3}$.

This near ring geometry is the expected one arising from the flat
limit of the  d1, d5 and kk dipole branes (see \KutasovZH\ for
discussion of string theory in this background, and \LarsenDH\ for
discussion of the boundary CFT.)  The corresponding conformal
field theory has $(4,0)$ supersymmetry, and $SU(2)$ R-symmetry. We
can think of this theory as the IR fixed point of an RG flow
starting from the $c= 6N_1 N_2$, $(4,4)$, CFT at the original AdS
boundary (the UV). Indeed, the IR central charge is strictly less
than that in the UV: $n_1 n_2 n_3 < N_1 N_2$.  In the UV we
started out in a nontrivial vacuum state which broke conformal
invariance, much like starting at some point on the Coulomb
branch. Some of the original degrees of freedom are massive in the
new vacuum, and the remaining massless ones give rise to the CFT
in the IR.

The definitions of $\psit$ and $\phit$ in the first line of \ea\
represent a  spectral flow by $\eta_L = -{1 \over n_3}$, and maps
the state to a state with $L_0 = \overline{L}_0 = c/24$ (same as
for the Ramond vacua) and $J_R =-J_L = {\Nb_1 \Nb_2 \over n_3}$.
Furthermore, the fermions are periodic up to the phases $e^{2\pi i
/n_3}$.  This agrees with what one expects for one of the ground
states of the tube string introduced earlier, consisting of
${\Nb_1 \Nb_2 \over n_3}$ component strings, each of length $2\pi
n_3$.

The redefinition $x_5 \rightarrow \xt_5$ corresponds to a more
novel redefinition of the superconformal generators.  It would be
very interesting to understand from the CFT point of view why
these new generators are the preferred ones in the IR; that is, to
understand the RG flow better.

\subsec{Case 2:~ $\delta \neq 0,~ \gamma=0$}

For nonzero $\delta$, after performing the same coordinate
transformation as in \ea\ we find the metric \eb\ plus the
additional term
\eqn\ed{ {C \delta \over \sqrt{q_1 q_2 q_3^4}}(dt+d\xt_5)^2~,}
corresponding to an extremal BTZ black hole.  In terms of the
Virasoro algebra of this near ring geometry, the level is
\eqn\eda{L_0 = \delta,}
which, combined with \ec\ and Cardy's formula $S= 2\pi \sqrt{{c
\over 6}L_0}$ yields the entropy formula \de\ for $\gamma=0$.

\subsec{Case 3:~ $\delta \neq 0,~ \gamma \neq0$}

We now write
\eqn\ee{\eqalign{ \psit & = \left(1 - {\gamma \over n_1 n_2
n_3}\right)\psi -{1 \over q_3} (x_5+t), \quad \phit  = \phi
+{\gamma \over n_1 n_2 n_3} \psi+{1 \over q_3}(x_5+t) \cr \xt_5 &
= q_3 \psi -t, \quad \cos (2\alpha)=x, \quad \yt  = -\sqrt{{q_1
q_2 q_3^2 \over 2 (Q_1 Q_2-q_3 C \delta)}} \sqrt{y}~, }}
to bring the leading form of the metric for $\yt \rightarrow
\infty$ to
\eqn\ef{\eqalign{ds^2 &= {\sqrt{q_1 q_2 q_3^2} \over \yt^2}(d\yt^2
-dt^2 +d\xt_5^2)+{C^2 \over (q_1 q_2 q_3^2)^{3/2}}(n_1 n_2 n_3
\delta -\gamma^2)(dt + d\xt_5)^2 \cr &~~+\sqrt{q_1 q_2
q_3^2}(d\alpha^2 + \sin^2 \alpha d\psit^2 + \cos^2 \alpha
d\phit^2)+ \sqrt{{q_2 \over q_1}}ds_{T^4}^2~.}}
If ${\gamma \over n_1 n_2}$ is an integer then the geometry is
that of AdS$_3  \times S^3 / Z_{n_3}\times T^4$, with the AdS$_3$
part being an extremal BTZ black hole, but more generally the
metric is not a product.   From \ee\ we see that there is an
additional transformation of the angular coordinates on top of the
spectral flow by $\eta_L = -{1 \over n_3}$.  The meaning of this
in terms of the original D1-D5 CFT is unclear to us.

\newsec{Discussion}

We have studied the black ring entropy from two points of view,
corresponding to the asymptotic D1-D5 CFT and the near ring
d1-d5-kk CFT.  In the former case, which is the appropriate one if
one wishes to achieve a common description of the black rings with
the two-charge supertubes and the BMPV black hole, we proposed
that the CFT effective string can be thought of as splitting into two
parts which describe separate properties of the black rings.   The
general entropy formula in this description matches the black ring
entropy after making a phenomenological assumption about the
lengths of the component strings.  Hopefully, a derivation of this assumption
from first principles will be supplied in the future.

By studying the geometry near the ring, we saw the emergence of
the d1-d5-kk system, and this lead to a simple understanding of
the general entropy formula. This analysis also showed that the
geometries dual to individual CFT microstates are singular,
despite having small curvature. The possibility of obtaining a
nonsingular solution via a more general supergravity Ansatz is
left open.

There are a number of  interesting aspects of these systems which
deserve further study. Perhaps the most intriguing one is the
relation between the length of the components of the CFT effective
string and the parameters in the geometry. In section 5 we have
argued that zero entropy geometries with nonzero $\gamma$ and
$\delta$ are dual to CFT microstates with component strings of
length given by \di, and have used this phenomenological
assumption to explain the entropy of all circular black rings.
However, there are many similar CFT microstates whose component
strings have a different length, yet their duals do not appear to
be among the $U(1) \times U(1)$ invariant solutions we have
considered. It is also unlikely that these microstates are dual to
non-circular rings (a naive time of travel calculation suggests
that the symmetry of the geometry does not change as one changes
the length of the component strings). One possibility is that
these represent multi-ring solutions
\refs{\GauntlettWH,\GauntlettQY}. It would be very interesting to
find the duals of these microstates, or, alternatively, explain
why only the microstates \di\ have geometric duals.  It is also
possible that these microstates simply do not exist when one takes
into account the deformations away from the orbifold point of the
CFT.

The $U(1) \times U(1)$ invariant black rings analyzed here appear
to be a  small subset of a much larger class of black rings of
arbitrary shape and charge densities, parameterized by seven
arbitrary functions \BW. The existence of this huge class of
solutions is also supported by the Born-Infeld analysis of three
charge supertubes of arbitrary shapes \BenaWT.  It is important to
try to map out the microscopic description of these more generic
geometries.


Another interesting avenue for future research is to better
understand the RG flow between the two CFT descriptions discussed
above. Since the near-ring metric only depends on the  dipole
charges and local branes densities of the black rings, but not on
their shapes, one expects the end point of the RG flows to be
universal, regardless of the shape of the ring. This very large
class of RG flows to the common IR fixed point deserves a more
thorough analysis.

\bigskip
\noindent {\bf Acknowledgments:} \medskip \noindent We would like to thank
Masaki Shigemori and Radu Roiban for discussions. The work of IB and PK is
supported in part by the NSF grants  PHY00-99590 and  PHY01-40151.

\listrefs
\end